\documentclass[acmsmall]{acmart}
\usepackage{verbatim}
\newcommand\mynote[1]{\textcolor{red}{#1}}

\usepackage[normalem]{ulem}
\useunder{\uline}{\ul}{}

\AtBeginDocument{%
  \providecommand\BibTeX{{%
    Bib\TeX}}}

\setcopyright{acmlicensed}
\copyrightyear{2024}
\acmYear{2024}
\acmDOI{XXXXXXX.XXXXXXX}

\acmConference[]{}{}{}    
\acmISBN{}

\usepackage{listings} 
\usepackage{glossaries} 
\usepackage{siunitx}

\newenvironment{packed_item}{
\begin{itemize}
  \setlength{\itemsep}{1pt}
  \setlength{\parskip}{0pt}
  \setlength{\parsep}{0pt}
}{\end{itemize}}
\newenvironment{packed_enum}{
\begin{enumerate}
  \setlength{\itemsep}{1pt}
  \setlength{\parskip}{0pt}
  \setlength{\parsep}{0pt}
}{\end{enumerate}}

  \newcounter{mnote}
  \newcommand{\mnote}[1]{\addtocounter{mnote}{1}
    \ensuremath{{}^{\bullet\arabic{mnote}}}
    \marginpar{\footnotesize\em\color{red}\ensuremath{\bullet\arabic{mnote}}#1}}
  \let\oldmarginpar\marginpar
    \renewcommand\marginpar[1]{\-\oldmarginpar[\raggedleft\footnotesize
#1]%
    {\raggedright\footnotesize #1}}

\newcommand\lbi{\begin{packed_item}}
\newcommand\lei{\end{packed_item}}

\newcommand{\fig}{Fig.~\ref}

\newcommand{\sect}{Sect.~\ref}

\newcommand{\tbl}{Table~\ref}

\newcommand{\verify}{\mynote{verify}}

\newcommand{\eqn}{Eq.~\ref}

\newcommand{\figsquick}[5]
{
\begin{figure}[ht]
\begin{center}
\includegraphics[width=#5]{#1}
\end{center}
\caption{
\mynote{(figshort:#2)}
#3\label{figshort:#2}.
\mynote{This figure was generated using the following notebook and dataset:}
\lstinline|#4|
}
\end{figure}
}

\newcommand{\figbig}[4]
{
  \figsquick{#1}{#2}{#3}{#4}
  {6in}
}

\newacronym{rag}{RAG}{retrieval augmented generation}
\newcommand\rag{\gls{rag}}
\newacronym{er}{ER}{endoplasmic reticulum}
\newcommand\er{\gls{er}}

\newacronym{llm}{LLM}{large language model}
\newcommand\llm{\gls{llm}}

\renewcommand\mynote[1]{}
\renewcommand\mnote[1]{}

\begin{document}

\title{Leveraging Retrieval-Augmented Generation and Large Language Models to Predict SERCA-Binding Protein Fragments from Cardiac Proteomics Data}

\author{Taylor A. Phillips} 
\author{Alejandro W. Huskey} 
\author{Patrick T. Huskey} 
\author{Seth L. Robia} 
\author{Peter M. Kekenes-Huskey}
\email{pkekeneshuskey@luc.edu}
\orcid{1234-5678-9012}
\authornotemark[1]
\affiliation{%
  \institution{Dept. Cell \& Molecular Physiology, Loyola University Chicago}
  \city{Maywood}
  \state{IL}
  \country{USA}
}
%
\renewcommand{\shortauthors}{Kekenes-Huskey}

\begin{abstract}
Large language models (\glsplural{llm}) have shown promise in various natural language processing tasks, including their application to proteomics data to classify protein fragments.
In this study, we curated a limited mass spectrometry dataset with 1000s of protein fragments, consisting of proteins that appear to be attached to the endoplasmic reticulum in cardiac cells, of which a fraction was cloned and characterized for their impact on SERCA, an ER calcium pump.
With this limited dataset, we sought to determine whether \glsplural{llm} could correctly predict whether a new protein fragment could bind SERCA, based only on its sequence and a few biophysical characteristics, such as hydrophobicity, determined from that sequence.
To do so, we generated random sequences based on cloned fragments, embedded the fragments into a \acrfull{rag} database to group them by similarity, then fine-tuned \llm\ prompts to predict whether a novel sequence could bind SERCA.
We benchmarked this approach using multiple open-source LLMs, namely the Meta/llama series, and embedding functions commonly available on the Huggingface repository.
We then assessed the generalizability of this approach in classifying novel protein fragments from mass spectrometry that were not initially cloned for functional characterization.
By further tuning the prompt to account for motifs, such as ER retention sequences, we improved the classification accuracy by and identified several proteins predicted to localize to the endoplasmic reticulum and bind SERCA, including Ribosomal Protein L2 and selenoprotein S.
Although our results were based on proteomics data from cardiac cells, our approach demonstrates the potential of LLMs in identifying novel protein interactions and functions with very limited proteomic data.

\end{abstract}

    
\received{26 Feb 2025}

\maketitle

\section{Introduction}
Conventional machine learning algorithms have provided powerful frameworks for analyzing proteomics data \cite{Beck2024}. 
However, these models often involve trade-offs between computational cost, data requirements, and accuracy.
In recent years, \llm s have gained traction as potentially inexpensive alternatives to analyze new data with minimal or no training required.
LLMs are a type of neural network trained on massive amounts of text data for tasks such as zero-shot sentence completion, where the model predicts the following words in a sequence without additional training. 
In particular, several LLMs have been specifically designed to relate protein sequences to structure, including ProtTrans~\cite{Elnaggar2022} and ESM~\cite{Lin2023}.
The field is rapidly evolving, including advances that incorporate physics engines and machine learning for de novo protein design and analysis~\cite{ghafarollahi2024protagents}. 
Progress has also been made in the integration of language models with geometric deep learning approaches to predict functionalities including protein-protein interfaces, docking, binding constants~\cite{Wu2023}, and coevolutionary patterns~\cite{Sgarbossa2023}.
These approaches exemplify the potential power and novelty of using LLMs to connect amino acid sequences to a broad range of biological phenomena.

Although \llm s demonstrate impressive zero-shot performance, their accuracy often benefits from additional relevant information.
Fine-tuning LLMs is a common strategy for tailoring them to specific tasks, but it typically requires significant data, computational resources, and expertise.
\Acrfull{rag} offers a more recent approach. 
It allows users to create a searchable local database that can be queried first to extract relevant information before consulting the LLM.
RAG databases are typically constructed by partitioning documents into smaller "chunks" that are then transformed into numerical vectors using embedding functions \cite{lewis2020retrieval}. 
These functions create vector representations that can be easily understood and processed by LLMs.  
Therefore, RAGs have the potential to enhance existing LLMs adapted for proteomics without the need for additional training. 
However, building and querying RAGs for optimal LLM performance can be a complex and laborious undertaking.

In this study, we used the paradigm \rag to predict the biophysical properties of protein fragments based on a larger data set curated from mass spectrometry-based proteomic analysis (unpublished).
The fragments originate from heart tissue, specifically within the \er\ of heart cells.
These protein fragments are believed to integrate into the ER membrane as transmembrane peptides, potentially binding to and disrupting the normal function of the sarcoplasmic reticulum calcium ATPase (SERCA), a calcium pump \cite{Bers2001}.
This manner of SERCA regulation is already well-established for other single-pass helical structures such as phospholamban (PLN)\cite{MacLennan2003} and dwarf open reading frame (DWORF)  \cite{Makarewich2018}.

A small, representative subset of the fragments was experimentally characterized in a separate study (Phillips Et al., MS in prep) to assess their impact on SERCA.
Our goal is to utilize these data to predict the ability of yet-uncharacterized fragments to modulate SERCA activity.

\section{Methods and Development}
\subsection{Data set}
\paragraph{Tissue}
Left ventricle tissue was collected from dilated cardiomyopathy (DCM) and non-failing human donor hearts. 
Triplicate samples were used for each group. 
The tissue was homogenized and fractionated by ultracentrifugation. 
The membrane fraction was retained from all samples and subjected to SDS-PAGE. 
Bands below 25 kD were excised from the gel. 
Gel pieces were subject to reduction with 10mM DTT and in-gel trypsin digestion. 
Protein was extracted from gel in preparation for mass spectrometry and resuspended in 3\% acetonitrile to a final concentration of 500 ng/uL.  

\paragraph{Proteomics} 
The excised bands were subjected to MS analysis via the Orbitrap Eclipse™ Tribrid™ Mass Spectrometer (FSN04-10000). 
The dataset contained approximately 8,900 fragments obtained through digestion of roughly 1800 proteins.
\mynote{(data available in ./data/fragments.csv)}
The protein fragments identified by MS were 15-30 amino acids in length. We selected 300 of which based on their 3-fold higher differential expression in DCM tissue relative to rejected transplants. We consulted UniProt [2] to select the subset comprising approximately 100 sequences that were confirmed to be localized in the ER and belong to a known transmembrane protein.
\mynote{upregulated.csv –> testupregulated.csv using annotateData.py }
Protein sequences corresponding to the ER-enriched TM data set were provided in FASTA format, where each of the twenty natural amino acids is represented by a single letter code. From this subset, we selected 10 fragments that bore a sequence similarity to a known SERCA inhibitor, phospholamban, for experimental characterization. 
\mynote{Raw data are in (poison peptide sequences.xlsx, 3/7/24)}

\paragraph{Functional characterization}
The procedure for characterization included cloning selected fragments into fluorescently tagged vector and transfection into HEK293T cells. 
Localization of protein fragments compared to SERCA was observed via structured illumination microscopy (SIM). 
Binding to SERCA was quantified via FRET microscopy and a KD value of 20 or under was used as a binary cutoff for high binding affinity. 
Consequently, the fragments were annotated with 1 or 0 for the activity attributes of binding SERCA. 
\mynote{see cloned.csv, which was generated from annotateData.py}

\paragraph{Data format and random sequence generation}
Since most of the 10 cloned sequences bound SERCA,  we generated randomized sequences derived from the training set to achieve a larger and more balanced dataset. 
For our generation strategy, we assumed that conservative amino acid replacements, such as aspartic acid (Asp) to glutamic acid (Glu), would result in `child' sequences that retained the activity of their `parent' sequences.  
We subjected each parent sequence to a single mutation event at a randomly chosen replacement site, generating a single sequence. 
Progeny from one generation would serve as parents for a successive mutation step.
This mutation and progeny generation process was iteratively applied to both parent and child sequences until a sufficient number of sequences ($\approx$ 250)  were generated to achieve a nearly balanced state for the dataset.
\mynote{see training.txt generated from featurizeData.py}

\subsection{\Acrfull{rag} + \Acrfull{llm}}
\paragraph{RAG} 
\label{meth:rag}
A \rag\ approach \cite{Salemi2024} was used in lieu of fine-tuning to augment pre-trained \llm s with information from our proteomics dataset.
Namely, a vector database was constructed with the dataset entries denoted as binding SERCA (1) or not binding SERCA (0) from the random mutagenesis strategy described above.
This was achieved by creating chunks (subsets) of the data with overlaps between adjacent chunks. 
These chunk sizes and overlaps were treated as hyperparameters. 
Embedding functions were selected from the Hugging Face repository, including examples such as chroma and faiss.
These embedding functions were used to map each input sequence to a corresponding vector with hundreds of floats. 
This was achieved using the \textsc{langchain-community} python libraries.

\paragraph{Clustering and feature expansion}
We partitioned the embedded data vectors onto a 2D plane using UMAP \cite{McInnes2018}, which is available as a python library.
UMAP aims to uniformly distribute data points along a manifold that reflects a `local' distance between neighboring points \cite{McInnes2018}, as defined by the points' embeddings. 
The points were then clustered using K-means via \textsc{scikit-learn} package, 
after which we retained a user-defined number of clusters.
In this fashion, fragments that share similar sequences and potentially SERCA-binding capacity, would reside in the same cluster.
A member of each cluster was further evaluated, namely the computation of its hydrophobic score, length, and charge density. 
We assumed that these physicochemical properties might be strong determinants of SERCA binding.
See \sect{equations} for their respective definitions.
These additional evaluations were appended to the input sequences that are later used for chain-of-thought prompting of the \llm s.
These attributes were also estimated for the novel sequences that were classified by the \llm s.

\paragraph{Human-guidance}
We also created a jupyter notebook (\textsc{manual\_select.ipynb}) to provide human-guided partitioning of the training data sets for supplemental annotation of the sequences.
Namely, we provide an option to assign a classification scheme to sequences, such as 0 for hydrophobic, 1 for polar, and 2 for charged.
The numerical values were used to color-code the sequences. 
The user is then asked to select sequences from a randomized set that are most similar to an input sequence. 
If the user is able to reliably classify the structures by eye, the classification scheme is used to add additional features to each input sequence or to fine-tune the \llm prompts, such as by reflecting observations of charge or hydrophobicity.

\paragraph{\llm\ prompting}
Prompts were compiled based on a chain of thought prompting \cite{Wei2022} using Input/Output pairs, followed by a query to predict the output of a novel sequence.
Namely, we provided several input/output pair examples with explanations of their sequences and properties.
As an example, the prompts could assume the form of the following: 
\lbi
\item \textsc{
I have a list of input/output pairs, where the input includes an amino acid sequence, a score, 
a length, and a hydrophobicity score, and the output is 1 or 0 if it binds a target protein.
Input: REGPPFISEGYAVRG0,0.118,16,0.2  Output: 0;
RNLQSEVDYGVKN1,0.111,14,0.3  Output: 1
Predict whether the following input has an output of 1 based first on its similarity to other sequences and their scores, make sure to explain each step:
RNKTEDLEATSEHFRYT1,0.115,18,0.4
}
\lei
Additional examples are included in \sect{supp:prompts} and our github repository.
These prompts are submitted to an \llm\ (see \tbl{tab:f1}) as a conversation chain, which is effectively a chatbot framework provided by \textsc{langchain}. 
Its responses were then parsed for predictions.
Performance was assessed for testing sets randomly generated without replacement from 10\% of the sequence data. 
F1, accuracy, and recall scores were computed with \textsc{scikit-learn} for several iterations of each embedding/LLM combination.

\section{Results and Discussion}
\subsection{Characterization of sequences derived from clone protein fragments}
Several fragments were identified by mass spectrometry that were ultimately cloned (see \tbl{tbl:cloned}). 
In \fig{figshort:structure} we show the fragments and their approximate locations in the encoded proteins, largely based on predicted structures from Alphafold \cite{Jumper2021}. 
For the transmembrane protein 50A (TMEM50A, O95807),  the fragment is a loop located between two transmembrane helices.
The functional role of TMEM50A is not well-established, though its bundle of four amphoteric helices suggests its localization to cell membranes.
For the small integral membrane protein 13 (SMIM13, P0DJ93), one fragment is positioned far from the transmembrane helix, while another is adjacent. 
The in vivo SMIM13 function has not yet been determined, but its resemblance to PLN is suggestive of its potential to interfere with SERCA function. 
For vesicle-associated membrane protein 3 (VAMP3, Q15836) and 
VAMP8 (Q9BV40),
the fragments are derived from the proteins' cytosol-facing N-termini (not shown).
VAMPs are well-known structures anchored in cell membranes that facilitate membrane fusion via SNARE complexes \cite{Hu2007}. 
For the ER membrane protein complex subunit 6 (EMC6, Q9BV81), the first fragment is adjacent to the transmembrane helix, while the second fragment is part of the helix's terminus. 
EMC6 was found to play an important role in autophagy  \cite{Li2013}.
For the transmembrane protein 14C (TMEM14C, Q9P0S9), the identified fragment spans a loop between two helices, based on NMR structures available from PDB Accession code 2LOS \cite{Klammt2012}.
Finally, in the SEC61 translocon subunit beta (SEC61B, S4R3B5), the fragment is located just after the transmembrane helix (not shown). 
This subunit plays an important role in protein transport across the ER membrane \cite{Rapoport2008}.
Hence, the fragments almost exclusively are found outside of the transmembrane helices. 

Since we have just a small number of cloned constructs available for model training, we expanded the dataset via generate constructs with one or more randomly-selected conservative replacements (such as D to E). 
This procedure was used to create approximately two hundred high-homology variants of the original cloned dataset. 
We also emphasized the generation of constructs that did not bind SERCA (namely EMC6 and TMEM14C), in order to generate a balanced dataset with comparable numbers of binders and non-binders.

\subsection{Analysis of the data set}
\fig{figshort:counts} summarizes characteristics of the SERCA-binding  and non-binding  protein fragments in our data set.
\fig{figshort:counts}A counts the number of each amino acid type in the binding  (blue) and nonbinding (orange) sequences.
The hydrophobic residues are presented initially (A-W), followed by the polar ones (Y-T), and finally, the charged amino acids (D, E, K, R). 
The binding sequences appear to have slightly higher numbers of polar and charged amino acids, but otherwise the counts are similar.
The lengths of the fragments in \fig{figshort:counts} B suggest that the longest sequences tend to bind to SERCA, but shorter sequences are intermixed.
Panel C suggests some tendency for binding sequences to have charge densities greater than 0.3, whereas non-binding sequences generally are less charged.
This trend is reasonable, given that the fragments were largely found adjacent to transmembrane peptides, and thus would be expected to favor polar residues in cytosolic and lumenal environments. 
Hence, although some small trends emerge, in general these features do not exhibit clear separation between the two groups, suggesting a non-trivial relationship with ER localization to bind SERCA. 
This aligns with our previous work~\cite{Sun2023}, where ultimately we used decision trees to effectively classify proteins with complex property-target relationships.  

\subsection{\Acrfull{rag}}
\subsubsection{Embedding}
Numerous machine learning methods, including \llm s exist for protein property classification based on sequence data \cite{Wasim2024,Wang2024,Krishna2024,Abramson2024}.
This study aimed to assess the potential of pre-trained \llm s to perform this task without fine-tuning. 
To achieve this, we employed a retrieval-augmented generation (RAG) approach \cite{ke2024development} that leverages embeddings to capture meaningful relationships between tokens from the fragments' FASTA sequences.
By encoding these sequences as numerical vectors, we enabled the retrieval of similar entries from a local RAG database.
Namely, the transformation from alphanumeric representations to vector embeddings ensures that sequences with shared features, e.g. sequence motifs, are positioned closely in the embedding space.
These retrieved sequences are then used to construct prompts for the LLM, allowing it to generate contextually relevant outputs.
In this fashion we bridge the gap between the static knowledge of pre-trained LLMs and the dynamic requirements of sequence-specific analyses of protein fragments. 

We analyzed the embeddings for the hundreds of sequences used for training in the compiled RAG database.
This was done by projecting them onto a 2D plane using UMAP (see \fig{figshort:cluster}A).
Similar sequences, such as those differing by a few amino acids, are found to colocalize with each other in a UMAP projection (see \fig{figshort:cluster}).
Moreover, sequences that were categorized as binding versus not binding appear to have nonoverlapping distributions. 
Namely, binding sequences (filled) appear to be well separated from non-binding sequences (outline) in this large dataset. 
Anecdotally, sequences within a given UMAP cluster seem to have similar characteristics, such as a high abundance of glutamine in the fragment from VAMP3 (\tbl{tbl:cloned}) that might also be similar to another TM protein like PLN (\tbl{tbl:posneg}).
This similarity becomes critical in the RAG step, for which we seek to match a trial sequence to a sequence with a known binding phenotype.

Using the separation observed in UMAP (\fig{figshort:cluster}B), we clustered the data using k-means to obtain a representative subset of the complete training set. 
\fig{figshort:cluster}C summarizes cluster sizes with bars shaded by Shannon entropy (\eqn{eqn:entropy}). 
The black bars represent clusters with an entropy approaching 0 (nearly homogeneous), while the white bars indicate an entropy approaching 1 (maximally mixed). 
Ideal partitioning and clustering would yield clusters with entropies close to 0, signifying groups with a single output value.
We randomly selected one data point from each k-means cluster (Figure~\fig{figshort:cluster}C) to represent local distributions in the UMAP projection. 
The sequences associated with these cluster representatives were scored using the matching results from the RAG query.

\subsubsection{Appending data}
Since other physical properties, such as charge density, would be expected to play a role in binding SERCA, each sequence was evaluated for charge density, fragment length, and hydrophobicity \mynote{, and dssp} (see Section~\ref{equations}). 
\mnote{Dssp - get alphafold structure, run dssp }
While these computationally inexpensive metrics provide limited biophysical detail, they serve as placeholders for more detailed calculations in future studies.
The scalar values for biophysical properties were appended to the input sequences, creating a combined FASTA format with sequence and numeric data, such as 
\begin{verbatim}
Input: RLQQTQNQVDEVIDIMRV0,0.116,20,0.3 Output:1; 
Input: REGPPFISEAAIRG0,0.139,15,0.3 Output: 0;
\end{verbatim}
This process was repeated for the test sequences. 
\mynote{includno centroid hydrophobicity \verify\eqn{eqn:hydrophobicity}}

\subsubsection{Chain of Thought Inquiry}
Finally, after appending the features to the sequences, we compiled them into the LLM prompts. 
Namely, the training data composed of the modified FASTA sequences and their SERCA binding phenotype were constructed as a chain of thought LLM prompt:
\begin{verbatim}
I have a list of input/output pairs, where the input includes an amino acid sequence, 
a score, length, and a hydrophobicity score, and the output is 1 or 0 if it binds 
a target protein.
Input: RLQQTQNQVDEVIDIMRV0,0.116,20,0.3 Output:1;
.....
Based on the provided inputs and outputs, predict the output of the following inputs:
RELVGDTGSQEGDHEPSGSETEEDTSSSPHRI0,0.098,33,0.3;
....
\end{verbatim}
We generally selected for 10 input/output pairs for the prompt and requested the LLM to predict the phenotype for test data. 
We included instructions in the prompt to format the responses into phrases that could be easily extracted for benchmarking.
The test sequences were not used for training and were scrubbed of the assigned SERCA binding phenotype.

\subsection{Benchmarking performance}
To benchmark our data, we randomly divided the data set into 90\% for training and 10\% for testing.
The predictions of the test data were compared with the true values for metrics such as F1 scores and accuracy. 
This process was repeated for 5 iterations.
\tbl{tab:f1} summarizes the F1 scores for various LLMs and embedding functions.
Scores approaching 1.0 represent strongly performing LLM/embedding combinations. 
Specifically, we found that the
llama3b:70b gave the best performance by the
F1 score ($0.96\pm0.06$) when used with
\textsc{e5-small-v2} embedding, 
1000 character chunking, and 0 character overlap. 
These parameters are summarized in \tbl{tbl:params}. 
The best performance was demonstrated with the larger LLM models (70b with 70 B parameters) and small (512 tokens) embeddings. 
We note that the different embeddings exhibited similar performances with the larger models, from which the rankings could moderately vary between benchmarking runs.

We further optimized hyperparameters, including the number of clusters and whether secondary attributes like charge density were appended to input/output pairs. 
Notably, omitting the ragScore yielded a modest improvement in F1-score (\textsc{2$^0$ attributes}, $0.95\pm0.064$). In contrast, not clustering sequences by UMAP resulted 
in the lowest score (\textsc{clusterSequences=False}, $0.66\pm0.256$). 
This suggests that distances between embeddings for a set of peptides were more important than their scoring method when queried. 
We did not vary chunk and overlap parameters, as these are typically relevant for large documents with smaller vector spaces. 
In our case, the entire sequence line was generally shorter than the length requirements of a given vector.

\subsection{Application to 'novel' fragments}
\subsubsection{Positive controls} 
We evaluated classification performance using fragments identified by mass spectrometry from proteins whose potential to bind to ER proteins is known. 
These included phospholamban (PLN), a known SERCA regulator \cite{Traaseth2008} 
\mnote{add me baCK - its truncated variant R14del \cite{vanderZwaag2012}, } 
and 
phospholemman (FYXD1), an FXYD protein that binds to the sodium=potassium exchanger ATPase on the plasma membrane \cite{Crambert2002}.  
Using our method, these TM proteins were predicted to bind SERCA (see \tbl{tbl:posneg}).

\subsubsection{Negative controls}
\label{sect:neg}
We also identified high-abundance fragments of cytosolic proteins in our ER fraction by mass
spectrometry, which were likely contaminants from sample preparation. 
These fragments were associated with components of the sarcomere such as Troponin C (TNNC1), the myosin light chains 1 and 3 (MYL1, MYL3) and actin (ACTA1).
Although the TM proteins were correctly predicted, the cytosolic proteins were initially also predicted to bind as well. 
We expanded the sequence to include up to twenty residues flanking the original fragment; for the TM proteins, the extensions contained sequences like CLILICLLLI that are associated with fragments that embed into membrane bilayers. 

In addition, after visually inspecting sequence properties such as charge density and amino acid type using the interactive notebook \textsc{manual\_select.ipynb}, we refined the prompt to account for features unique to positive controls, such as the presence of consecutive hydrophobic residues.
Interestingly, upon asking the LLM to suggest additional sequence patterns that differentiate positive from negative controls, it proposed the sequence of R followed by K or R.
We find the recommendation plausible, given the established role of RR in the localization of phospholamban in the endoplasmic reticulum \cite{Sharma2010} and the motifs RXR and KKXX for the localization of other peptides \cite{Shikano2003}.
Altogether, the refined attributes are consistent with the insertion of the sequence into the hydrophobic interior of a lipid bilayer and the stability of electrostatic interactions with negatively charged phosphate groups on the membrane surface. 
To reflect these observations, we appended to our prompt the statement
\begin{verbatim}
We observe that sequences with an output of 1
have several consecutive hydrophobic residues and the sequence RR.   
\end{verbatim}
After tuning the prompt, the negative controls were correctly predicted as not binding, while the positive controls maintained their correct predictions. 
\mnote{how does it make its descision - have Patrick ask for additional rationale }

\subsubsection{Full set of fragments}
Lastly, we applied the method and \textsc{llama3:70b/gte-small} combination to the remaining upregulated fragments in the data set (see \tbl{tbl:sequence_matches}).
We specifically narrowed the set to those that were determined to reside in the ER membrane, based on Uniprot annotations. 
Examples of hits predicted to bind include
cytochrome P450 2J2 (P51589), ribosomal protein L26 (P62854), 
3-beta-hydroxysteroid-Delta (8),Delta (7)-isomerase (Q15125), 
Selenoprotein S (Q9PQEV) and
dolichol-phosphate mannosyltransferase 1 (Q9P2X0).
The predicted binders all serve established roles in the ER.
Cytochrome P450 2J2, a key enzyme expressed in cardiac cells and bound to
the endoplasmic reticulum membrane, plays a crucial role in
detoxification and metabolism processes \cite{Zhao2021}.
Furthermore, ribosomal protein L26, a component of the large ribosomal subunit, is located in the rough ER and is involved in protein synthesis \cite{Scavone2023}.  
3-beta-hydroxysteroid-Delta(8),Delta(7)-isomerase, a transmembrane protein embedded in the ER
membrane and involved in cholesterol biosynthesis, specifically
facilitates a critical step in cholesterol formation\cite{SilventePoirot2012}. 
Other proteins with different functions include Selenoprotein S, a membrane protein that incorporates selenium into its active site, thus protecting cells from oxidative damage\cite{Liu2013},
and Dolichol-phosphate mannosyltransferase 1, an ER-associated enzyme involved in N-glycan biosynthesis and protein quality control\cite{Maeda2000}.
Importantly, these predictions show that our strategy for prompting an LLM with sequence-specific training data generalizes to novel fragments not found in our training set.

\subsection{Limitations} 
Our approach has several limitations that warrant further refinement. 
The most significant one is that our small data set size likely challenges the generalization of this \llm approach to diverse protein sequences that have little homology to the training data. 
Hence, a bias is introduced into the \llm approach from the input/output training sequences that we generated from our cloning data.
Naturally, this will likely miss functionally relevant sequences that were not part of cloned peptides, although this concern is mitigated somewhat by our successful testing against control sequences.
It is also possible that our approach may have missed important sequence motifs within transmembrane regions of the proteins that determine SERCA binding.
This is because the broad set of protein fragments identified by mass spectrometry tended to be from fragments that are exposed to the cell cytosol, whereas we generally did not observe protein domains likely to be embedded in lipid bilayers.

The current implementation considers a basic feature set for protein fragments. 
Including more details such as the predicted secondary structure \cite{Kabsch1983} and annotating known functional motifs, such as sequences rich in asparigine, leucine, and isoleucine that typically bind to SERCA \cite{Alford2020}, could potentially improve the performance of the LLM approach. 
Annotation of sequences that have ER retention motifs like RR may also be helpful in this regard. 
Lastly, expanding the data by incorporating external APIs for structure prediction and protein-protein interactions could be beneficial \cite{Abramson2024}. 

In terms of the \llm\ models used, there are many limitations to consider. 
One of these is the limited size of each \llm's context window, which restricts the extent of chain-of-thought prompting that can be done. 
In addition, the quality of the prompt used for the \llm s was an important determinant of obtaining sensible responses. 
For example, our experiments with Microsoft's phi3 medium \llm\ model were largely unsuccessful in generating interpretable responses, despite our efforts to guide its interpretation of our data and the formatting of the replies. 
Along these lines, it is expected that each new model that is used in our framework will need some prompt tuning to obtain coherent responses and accurate predictions.
Even after tuning the models, each model is intrinsically non-deterministic in its responses, which can add variability to its classification of trial sequences. 
However, questions that prompt the LLM to explain its reasoning might provide valuable insight toward optimizing prompts. 
Moreover, combining LLMs with human-interpretable models~\cite{Bordt2024} could be a promising strategy to guide this process.
Lastly and on a pragmatic note, we used \textsc{ollama} and \text{langchain} to implement our approach; these utilities and libraries are in a rapid stage of development, thus some features become obsolete with time and require some code updating.

\section{Conclusions}
This study explored the use of LLMs to predict the in vivo behavior of peptide fragments based on their sequences. 
A RAG database was created using peptide sequences with experimental characterization of SERCA binding. 
This database was used to train an LLM to predict the SERCA binding potential of the control and new peptide fragments. 
The results demonstrate the potential of LLMs, coupled with RAG, to accurately predict SERCA binding based on peptide sequence alone. 
This approach, although trained on protein fragments isolated from cardiac cells, is expected to generalize to other proteomic data comprising primary sequences. 
Further investigation into diverse protein datasets and additional sequence features may enhance the accuracy and applicability of this LLM-based approach.
Using these data, it may be feasible to extend the binary classification of SERCA binding to instead predict fragment/SERCA binding affinities.
\clearpage
\newpage
\section{Tables}

\begin{table}[h!]
\caption{\textbf{Cloned sequences} and representative fragments.
One additional fragment each for P0DJ93, Q15836, Q9BV81, and Q9P0S9 was identified, but not included in this table. 
These fragments are available in the github repository.  
Structures of representative examples are in \fig{figshort:structure}
}
\label{tbl:cloned}
\centering
\begin{tabular}{lllll}
\toprule
\textbf{Protein} & \textbf{Name} & \textbf{Sequence} & \textbf{Binds?} & \textbf{Accession \#} \\
\midrule
TMEM50A & Transmembrane protein 50A & RGDSYSEGCLGQTGARI & Y & O95807\\
SMIM13 & Small integral membrane prot. 13 & RRAPADEGHRPLT & Y & P0DJ93\\
VAMP3 & Vesicle-associated membrane prot. 3 & RRLQQTQNQVDEVVDIMRV & Y &  Q15836 \\
VAMP8 & Vesicle-associated membrane prot. 8 & RNLQSEVEGVKN & Y & Q9BV40 \\
SEC61B & SEC61 translocon subunit beta & RFYTEDSPGLKV & Y &S4R3B5\\
EMC6 & ER membrane prot. complex subunit 6 & REGPPFISEAAVRG & N &Q9BV81 \\
TMEM14C & Transmembrane protein 14C & KVGVSMFNRPH- & N &Q9P0S9 \\
\bottomrule
\end{tabular}
\end{table}

\begin{table}[h!]
\caption{\textbf{Performance} (F1 scores) with respect to embedding functions and LLMs available on HuggingFace.
The reported standard deviation is based on N=5 trials.
Below these trials, we show different model configurations, including appended features, numbers of clusters, clustering versus random, and I/O pairs.
$^*$ noragscore.
These were computed using a high-performing model/embedding pair as a reference case (e.g. llama3:70b/e5-small-v2).
\textsc{clusterSequences=False} entails randomly selecting points as opposed to choosing from clustering of rag scores.
\textsc{iopairPrompt=NoFasta} signifies omitting fasta sequence data and relying solely on attributes for chain of thought prompting. 
\mynote{See comments above for execution} 
}
  \label{tab:f1}
\begin{tabular}{@{}llllll@{}}
\toprule
\textbf{Embeddings} & \multicolumn{4}{l}{\textbf{LLM models}} \\ \midrule
                          &{\ul llama2:70b}& {\ul llama3:8b} & {\ul llama3:70b} & {\ul gemma2:27b} \\ 
\textsc{e5-small-v2}        & $0.63 \pm 0.03$ & $0.88 \pm 0.05$ & $\mathbf{0.96\pm0.06}$ & $0.85 \pm 0.12$\\  
\textsc{bge-small-en-v1.5}  & $0.61 \pm 0.10$ & $0.59 \pm 0.34$ & $        0.92\pm0.07 $ & $0.96 \pm 0.05$\\ 
\textsc{gte-small}          & $0.74 \pm 0.16$ & $0.48 \pm 0.33$ & $        0.94\pm0.05 $ & $0.88 \pm 0.07$\\  
\textsc{protein-matryoshka} & $0.90 \pm 0.13$  & $0.77 \pm 0.29$ & $        0.86\pm 0.11$ & $0.88 \pm 0.12 $ \\  
\textsc{snowflake-arctic-}  & $0.47 \pm 0.35$ & $0.74 \pm 0.30$ & $        0.99 \pm0.02$ & $0.91 \pm 0.11$ \\  
\hspace{5pt}\textsc{embed-xs}\\
\bottomrule
\textbf{Features} & \multicolumn{4}{l}{\textbf{LLM model }} \\ 
\midrule
\textbf{reference}       	 &      & & $\mathbf{0.96 \pm 0.06}$\\        
ragScore only				 &      & & $0.89 \pm 0.05$\\        
2$^o$ attributes$^*$         &      & & $0.95 \pm 0.06$\\        
no features                  &      & & $0.96 \pm 0.03$\\        
nClusters=10                 &      & & $0.95 \pm 0.05$\\        
clusterSequences=False       &      & & $0.66 \pm 0.26$\\        
iopairPrompt=NoFasta         &      & & $0.86 \pm 0.04$\\        
\bottomrule
\end{tabular}
\end{table}

\begin{table}[h!]
\caption{\textbf{Positive and negative controls}
Fragments were selected from the MS data for evaluation by the best performing LLM parameterization. 
Fragments were derived from two transmembrane protein peptides (PLN and FXYD1) and several sarcomeric proteins (TNNC1, MYL1, MYL3, ACTA1).
TNNC1 and MYL3 had several fragments.
Each fragment was flanked by several amino acids on each terminus. 
Prompt is in \sect{sect:neg}.
\mynote{See comments below for execution} 
\mynote{Accession numbers are commented out below}
}
\label{tbl:posneg}
\centering
\begin{tabular}{llllc}
\toprule
\textbf{Protein} & \textbf{Name} & \textbf{Fragment} & \textbf{Binding}\\ &&&\textbf{prediction}\\ 
\midrule
	PLN &  Cardiac phospholamban & RRASTIEMPQQARQ & Y \\ 
	FXYD1 &  Phospholemman & RTGEPDEEEGTFRS & Y  \\ 
	TNNC1 &  Troponin C, slow skel./card. & KIMLQATGETITEDDIEELMKD & Y  \\ 
	MYL1 &  Myosin light chain 1, skeletal & KMKEEEVEALMAGQEDSNGCINYEAFVKH & N  \\ 
	MYL3 &  Myosin light chain 3 & RLTEDEVEKLMAGQEDSNGCINYEAFVKH  & N \\ 
	ACTA1 &  Actin, alpha skeletal  & KDLYANNVMSGGTTMYPGIADRM & N \\ 
\bottomrule
\end{tabular}
\end{table}

\begin{table}[h!]
\caption{\textbf{Other upregulated candidates} Example predicted binders identified from set of upregulated fragments, using the default LLM parameterization.
Selected one fragment from each hit. 
\mynote{See comments below for execution} 
}
\label{tbl:sequence_matches}
\centering
\begin{tabular}{llllc}
\toprule
\textbf{Protein} & \textbf{Name} & \textbf{Fragment} & \textbf{Binds?} \\ 
\midrule
CYP2J2 & Cytochrome P450 2J2 & KGTMILTNLTALHRD & Y \\
RPS26 & Ribosomal Protein L26 & RDISEASVFDAYVLPKL & Y \\
EBP & 3-beta-hydroxysteroid-Delta(8), & KHLTHAQSTLDAKA & Y \\
&Delta(7)-isomerase\\
MMGT1 & EMC5 & KGLVGIGLFALAHAAFSAAQHRS & N \\
SELENOS & Selenoprotein S & RAAAAVEPDVVVKR & Y \\
DPM3 & Dolichol-phosphate  & RVATFHDCEDAARE & Y \\
&mannosyltransferase 3 \\
EMC6 & ER membrane protein 6 & RGNAAVLDYCRT & N \\
\bottomrule
\end{tabular}
\end{table}
\clearpage
\newpage

\section{Figures}
\figbig{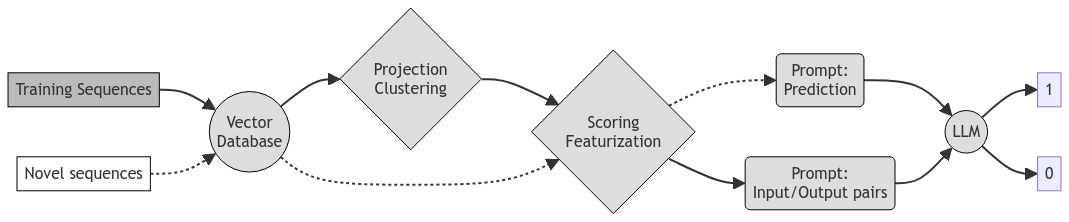}{schematic}{Our \Acrfull{rag}/\acrfull{llm} schematic for analyzing proteomics data in this study}{}

\figbig{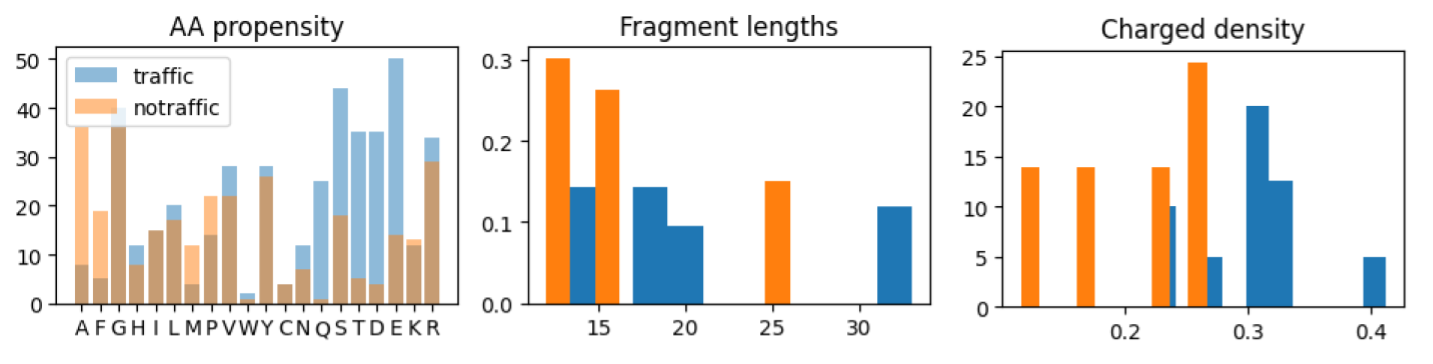}{counts}{\textbf{Protein fragment characteristics} 
A. Numbers of amino acids in sequences that bind (traffic, blue) or do not bind (notraffic, orange) SERCA.
B. Histogram of peptide lengths as a percentage of the total pool.
C. Histogram of peptide absolute charge densities in absolute numbers.
\mynote{replace w bind/not bind}
}
{fragments.ipynb}

\figbig{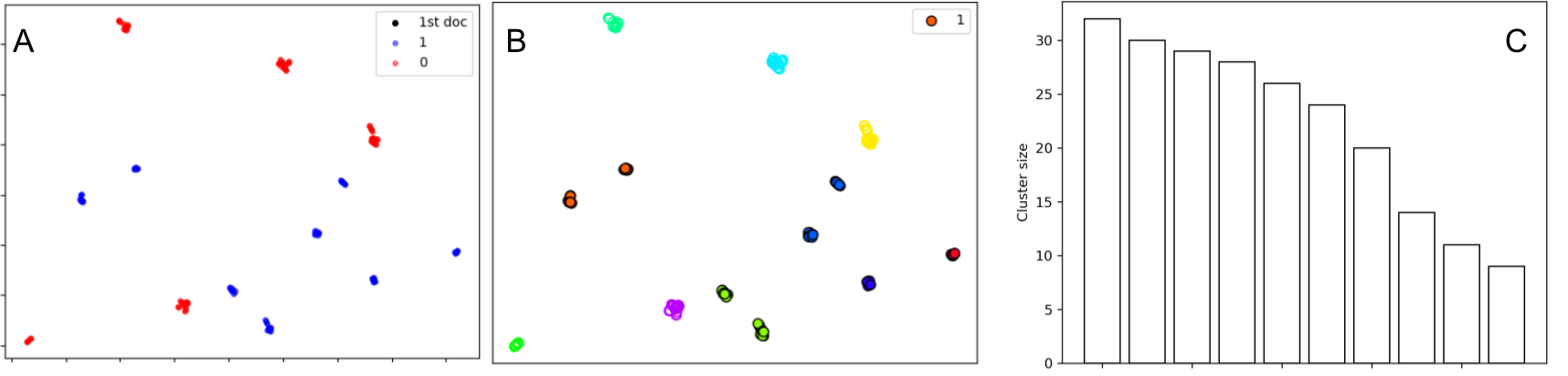}{cluster}{
\textbf{Clustering results}
UMAP results for a representative training set labeled according to 
A. binding (1, blue)/nonbinding (0, red) status.
B. Cluster assignments from A. by color. Points with black edges denote members of the binding class.
X- and Y- axes correspond to the projected UMAP coordinates.
C. Rank order of selected clusters by size.
}{
outputted by umapPrompter.py
}
\clearpage
\newpage

\figbig{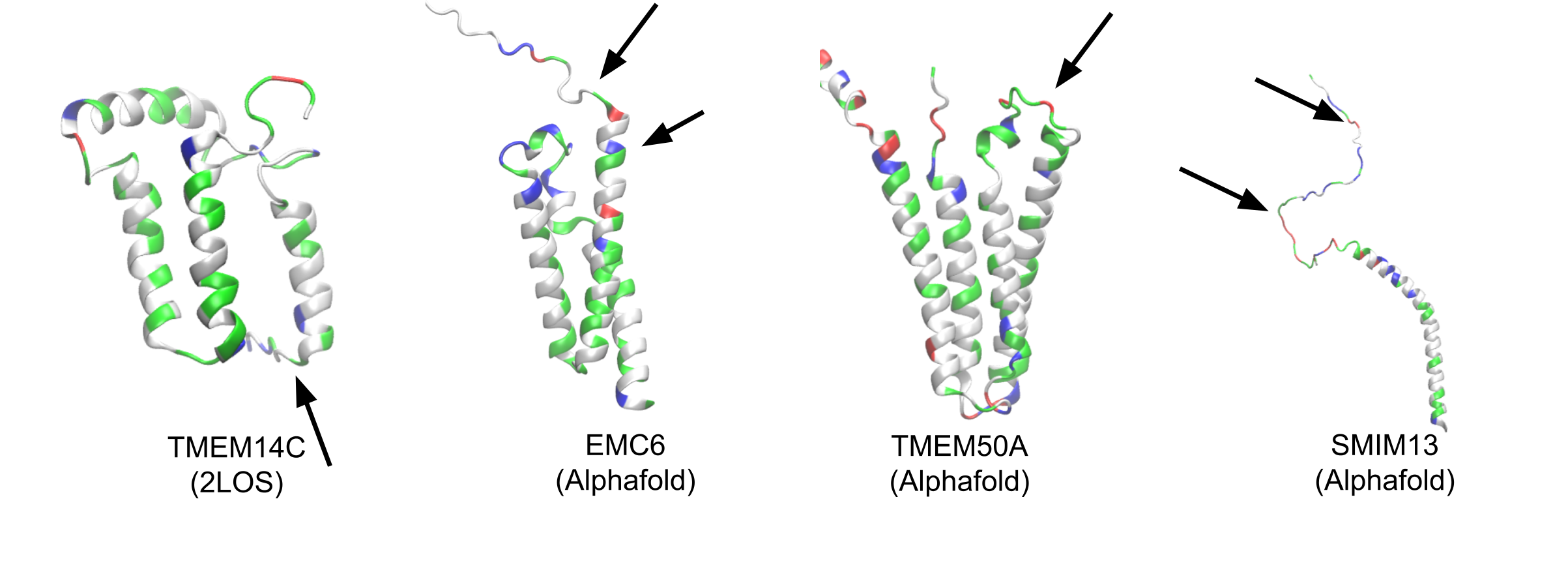}{structure}
{\textbf{Protein structures with cleaved fragments}
TMEM14C (PDB code 2LOS  \cite{Kabsch1983}), 
EMC6 (Alphafold accession \cite{Jumper2021}),
TMEM50A (Alphafold accession O95807), and 
SMIM13 (Alphafold ccession P0DJ93) are shown as ribbons colored according to amino acid type 
(polar, green; apolar, white; negatively-charged, red; positively-charged, blue).
Fragments identified via MS are identified with black arrows. 
}{on uniprot}

\begin{acks}
We are grateful for the Jonathan Kirk lab providing mass spectrometry analyses of the fragments.
We acknowledge generous support from the NIH in the form of an R35 grant to PMKH.
\end{acks}

\bibliographystyle{ACM-Reference-Format}
\bibliography{Acm}


\begin{thebibliography}{36}


\ifx \showCODEN    \undefined \def \showCODEN     #1{\unskip}     \fi
\ifx \showDOI      \undefined \def \showDOI       #1{#1}\fi
\ifx \showISBNx    \undefined \def \showISBNx     #1{\unskip}     \fi
\ifx \showISBNxiii \undefined \def \showISBNxiii  #1{\unskip}     \fi
\ifx \showISSN     \undefined \def \showISSN      #1{\unskip}     \fi
\ifx \showLCCN     \undefined \def \showLCCN      #1{\unskip}     \fi
\ifx \shownote     \undefined \def \shownote      #1{#1}          \fi
\ifx \showarticletitle \undefined \def \showarticletitle #1{#1}   \fi
\ifx \showURL      \undefined \def \showURL       {\relax}        \fi
\providecommand\bibfield[2]{#2}
\providecommand\bibinfo[2]{#2}
\providecommand\natexlab[1]{#1}
\providecommand\showeprint[2][]{arXiv:#2}

\bibitem[Abramson et~al\mbox{.}(2024)]%
        {Abramson2024}
\bibfield{author}{\bibinfo{person}{Josh Abramson}, \bibinfo{person}{Jonas
  Adler}, \bibinfo{person}{Jack Dunger}, \bibinfo{person}{Richard Evans},
  \bibinfo{person}{Tim Green}, \bibinfo{person}{Alexander Pritzel},
  \bibinfo{person}{Olaf Ronneberger}, \bibinfo{person}{Lindsay Willmore},
  \bibinfo{person}{Andrew~J. Ballard}, \bibinfo{person}{Joshua Bambrick},
  \bibinfo{person}{Sebastian~W. Bodenstein}, \bibinfo{person}{David~A. Evans},
  \bibinfo{person}{Chia-Chun Hung}, \bibinfo{person}{Michael O’Neill},
  \bibinfo{person}{David Reiman}, \bibinfo{person}{Kathryn Tunyasuvunakool},
  \bibinfo{person}{Zachary Wu}, \bibinfo{person}{Akvilė Žemgulytė},
  \bibinfo{person}{Eirini Arvaniti}, \bibinfo{person}{Charles Beattie},
  \bibinfo{person}{Ottavia Bertolli}, \bibinfo{person}{Alex Bridgland},
  \bibinfo{person}{Alexey Cherepanov}, \bibinfo{person}{Miles Congreve},
  \bibinfo{person}{Alexander~I. Cowen-Rivers}, \bibinfo{person}{Andrew Cowie},
  \bibinfo{person}{Michael Figurnov}, \bibinfo{person}{Fabian~B. Fuchs},
  \bibinfo{person}{Hannah Gladman}, \bibinfo{person}{Rishub Jain},
  \bibinfo{person}{Yousuf~A. Khan}, \bibinfo{person}{Caroline M.~R. Low},
  \bibinfo{person}{Kuba Perlin}, \bibinfo{person}{Anna Potapenko},
  \bibinfo{person}{Pascal Savy}, \bibinfo{person}{Sukhdeep Singh},
  \bibinfo{person}{Adrian Stecula}, \bibinfo{person}{Ashok Thillaisundaram},
  \bibinfo{person}{Catherine Tong}, \bibinfo{person}{Sergei Yakneen},
  \bibinfo{person}{Ellen~D. Zhong}, \bibinfo{person}{Michal Zielinski},
  \bibinfo{person}{Augustin Žídek}, \bibinfo{person}{Victor Bapst},
  \bibinfo{person}{Pushmeet Kohli}, \bibinfo{person}{Max Jaderberg},
  \bibinfo{person}{Demis Hassabis}, {and} \bibinfo{person}{John~M. Jumper}.}
  \bibinfo{year}{2024}\natexlab{}.
\newblock \showarticletitle{Accurate structure prediction of biomolecular
  interactions with AlphaFold 3}.
\newblock \bibinfo{journal}{\emph{Nature}} \bibinfo{volume}{630},
  \bibinfo{number}{8016} (\bibinfo{date}{May} \bibinfo{year}{2024}),
  \bibinfo{pages}{493–500}.
\newblock
\urldef\tempurl%
\url{https://doi.org/10.1038/s41586-024-07487-w}
\showDOI{\tempurl}


\bibitem[Alford et~al\mbox{.}(2020)]%
        {Alford2020}
\bibfield{author}{\bibinfo{person}{Rebecca~F. Alford}, \bibinfo{person}{Nikolai
  Smolin}, \bibinfo{person}{Howard~S. Young}, \bibinfo{person}{Jeffrey~J.
  Gray}, {and} \bibinfo{person}{Seth~L. Robia}.}
  \bibinfo{year}{2020}\natexlab{}.
\newblock \showarticletitle{Protein docking and steered molecular dynamics
  suggest alternative phospholamban-binding sites on the SERCA calcium
  transporter}.
\newblock \bibinfo{journal}{\emph{Journal of Biological Chemistry}}
  \bibinfo{volume}{295}, \bibinfo{number}{32} (\bibinfo{date}{August}
  \bibinfo{year}{2020}), \bibinfo{pages}{11262–11274}.
\newblock
\urldef\tempurl%
\url{https://doi.org/10.1074/jbc.ra120.012948}
\showDOI{\tempurl}


\bibitem[Beck et~al\mbox{.}(2024)]%
        {Beck2024}
\bibfield{author}{\bibinfo{person}{Armen~G. Beck}, \bibinfo{person}{Matthew
  Muhoberac}, \bibinfo{person}{Caitlin~E. Randolph}, \bibinfo{person}{Connor~H.
  Beveridge}, \bibinfo{person}{Prageeth~R. Wijewardhane},
  \bibinfo{person}{Hilkka~I. Kenttämaa}, {and} \bibinfo{person}{Gaurav
  Chopra}.} \bibinfo{year}{2024}\natexlab{}.
\newblock \showarticletitle{Recent Developments in Machine Learning for Mass
  Spectrometry}.
\newblock \bibinfo{journal}{\emph{ACS Measurement Science Au}}
  \bibinfo{volume}{4}, \bibinfo{number}{3} (\bibinfo{date}{February}
  \bibinfo{year}{2024}), \bibinfo{pages}{233–246}.
\newblock
\urldef\tempurl%
\url{https://doi.org/10.1021/acsmeasuresciau.3c00060}
\showDOI{\tempurl}


\bibitem[Bers(2001)]%
        {Bers2001}
\bibfield{author}{\bibinfo{person}{Donald~M Bers}.}
  \bibinfo{year}{2001}\natexlab{}.
\newblock \bibinfo{booktitle}{\emph{Excitation-Contraction Coupling and Cardiac
  Contractile Force}}. Vol.~\bibinfo{volume}{1}.
\newblock \bibinfo{publisher}{Kluwer Academic Publishers}. 427 pages.
\newblock
\showISBNx{9780792371588}
\showISSN{01669842}
\urldef\tempurl%
\url{https://doi.org/10.1007/978-94-010-0658-3_8}
\showDOI{\tempurl}


\bibitem[Bordt et~al\mbox{.}(2024)]%
        {Bordt2024}
\bibfield{author}{\bibinfo{person}{Sebastian Bordt}, \bibinfo{person}{Ben
  Lengerich}, \bibinfo{person}{Harsha Nori}, {and} \bibinfo{person}{Rich
  Caruana}.} \bibinfo{year}{2024}\natexlab{}.
\newblock \bibinfo{title}{Data Science with LLMs and Interpretable Models}.
\newblock
\newblock
\urldef\tempurl%
\url{https://arxiv.org/abs/2402.14474}
\showURL{%
\tempurl}


\bibitem[Crambert et~al\mbox{.}(2002)]%
        {Crambert2002}
\bibfield{author}{\bibinfo{person}{Gilles Crambert}, \bibinfo{person}{Maria
  Füzesi}, \bibinfo{person}{Haim Garty}, \bibinfo{person}{Steven Karlish},
  {and} \bibinfo{person}{Käthi Geering}.} \bibinfo{year}{2002}\natexlab{}.
\newblock \showarticletitle{Phospholemman (FXYD1) associates with Na,K-ATPase
  and regulates its transport properties}.
\newblock \bibinfo{journal}{\emph{Proceedings of the National Academy of
  Sciences}} \bibinfo{volume}{99}, \bibinfo{number}{17} (\bibinfo{date}{August}
  \bibinfo{year}{2002}), \bibinfo{pages}{11476–11481}.
\newblock
\urldef\tempurl%
\url{https://doi.org/10.1073/pnas.182267299}
\showDOI{\tempurl}


\bibitem[Elnaggar et~al\mbox{.}(2022)]%
        {Elnaggar2022}
\bibfield{author}{\bibinfo{person}{Ahmed Elnaggar}, \bibinfo{person}{Michael
  Heinzinger}, \bibinfo{person}{Christian Dallago}, \bibinfo{person}{Ghalia
  Rehawi}, \bibinfo{person}{Yu Wang}, \bibinfo{person}{Llion Jones},
  \bibinfo{person}{Tom Gibbs}, \bibinfo{person}{Tamas Feher},
  \bibinfo{person}{Christoph Angerer}, \bibinfo{person}{Martin Steinegger},
  \bibinfo{person}{Debsindhu Bhowmik}, {and} \bibinfo{person}{Burkhard Rost}.}
  \bibinfo{year}{2022}\natexlab{}.
\newblock \showarticletitle{{ProtTrans}: Toward understanding the language of
  life through self-supervised learning}.
\newblock \bibinfo{journal}{\emph{IEEE Trans. Pattern Anal. Mach. Intell.}}
  \bibinfo{volume}{44}, \bibinfo{number}{10} (\bibinfo{date}{Oct.}
  \bibinfo{year}{2022}), \bibinfo{pages}{7112--7127}.
\newblock


\bibitem[Ghafarollahi and Buehler(2024)]%
        {ghafarollahi2024protagents}
\bibfield{author}{\bibinfo{person}{A. Ghafarollahi} {and}
  \bibinfo{person}{M.~J. Buehler}.} \bibinfo{year}{2024}\natexlab{}.
\newblock \bibinfo{title}{ProtAgents: Protein discovery via large language
  model multi-agent collaborations combining physics and machine learning}.
\newblock
\newblock
\showeprint[arxiv]{2402.04268}~[cond-mat.soft]


\bibitem[Hu et~al\mbox{.}(2007)]%
        {Hu2007}
\bibfield{author}{\bibinfo{person}{Chuan Hu}, \bibinfo{person}{Deborah Hardee},
  {and} \bibinfo{person}{Fred Minnear}.} \bibinfo{year}{2007}\natexlab{}.
\newblock \showarticletitle{Membrane fusion by VAMP3 and plasma membrane
  t-SNAREs}.
\newblock \bibinfo{journal}{\emph{Experimental Cell Research}}
  \bibinfo{volume}{313}, \bibinfo{number}{15} (\bibinfo{date}{September}
  \bibinfo{year}{2007}), \bibinfo{pages}{3198–3209}.
\newblock
\urldef\tempurl%
\url{https://doi.org/10.1016/j.yexcr.2007.06.008}
\showDOI{\tempurl}


\bibitem[Jumper et~al\mbox{.}(2021)]%
        {Jumper2021}
\bibfield{author}{\bibinfo{person}{John Jumper}, \bibinfo{person}{Richard
  Evans}, \bibinfo{person}{Alexander Pritzel}, \bibinfo{person}{Tim Green},
  \bibinfo{person}{Michael Figurnov}, \bibinfo{person}{Olaf Ronneberger},
  \bibinfo{person}{Kathryn Tunyasuvunakool}, \bibinfo{person}{Russ Bates},
  \bibinfo{person}{Augustin Žídek}, \bibinfo{person}{Anna Potapenko},
  \bibinfo{person}{Alex Bridgland}, \bibinfo{person}{Clemens Meyer},
  \bibinfo{person}{Simon A.~A. Kohl}, \bibinfo{person}{Andrew~J. Ballard},
  \bibinfo{person}{Andrew Cowie}, \bibinfo{person}{Bernardino Romera-Paredes},
  \bibinfo{person}{Stanislav Nikolov}, \bibinfo{person}{Rishub Jain},
  \bibinfo{person}{Jonas Adler}, \bibinfo{person}{Trevor Back},
  \bibinfo{person}{Stig Petersen}, \bibinfo{person}{David Reiman},
  \bibinfo{person}{Ellen Clancy}, \bibinfo{person}{Michal Zielinski},
  \bibinfo{person}{Martin Steinegger}, \bibinfo{person}{Michalina Pacholska},
  \bibinfo{person}{Tamas Berghammer}, \bibinfo{person}{Sebastian Bodenstein},
  \bibinfo{person}{David Silver}, \bibinfo{person}{Oriol Vinyals},
  \bibinfo{person}{Andrew~W. Senior}, \bibinfo{person}{Koray Kavukcuoglu},
  \bibinfo{person}{Pushmeet Kohli}, {and} \bibinfo{person}{Demis Hassabis}.}
  \bibinfo{year}{2021}\natexlab{}.
\newblock \showarticletitle{Highly accurate protein structure prediction with
  AlphaFold}.
\newblock \bibinfo{journal}{\emph{Nature}} \bibinfo{volume}{596},
  \bibinfo{number}{7873} (\bibinfo{date}{July} \bibinfo{year}{2021}),
  \bibinfo{pages}{583–589}.
\newblock
\urldef\tempurl%
\url{https://doi.org/10.1038/s41586-021-03819-2}
\showDOI{\tempurl}


\bibitem[Kabsch and Sander(1983)]%
        {Kabsch1983}
\bibfield{author}{\bibinfo{person}{Wolfgang Kabsch} {and}
  \bibinfo{person}{Christian Sander}.} \bibinfo{year}{1983}\natexlab{}.
\newblock \showarticletitle{Dictionary of protein secondary structure: Pattern
  recognition of hydrogen‐bonded and geometrical features}.
\newblock \bibinfo{journal}{\emph{Biopolymers}} \bibinfo{volume}{22},
  \bibinfo{number}{12} (\bibinfo{date}{December} \bibinfo{year}{1983}),
  \bibinfo{pages}{2577–2637}.
\newblock
\urldef\tempurl%
\url{https://doi.org/10.1002/bip.360221211}
\showDOI{\tempurl}


\bibitem[Ke et~al\mbox{.}(2024)]%
        {ke2024development}
\bibfield{author}{\bibinfo{person}{YuHe Ke}, \bibinfo{person}{Liyuan Jin},
  \bibinfo{person}{Kabilan Elangovan}, \bibinfo{person}{Hairil~Rizal Abdullah},
  \bibinfo{person}{Nan Liu}, \bibinfo{person}{Alex Tiong~Heng Sia},
  \bibinfo{person}{Chai~Rick Soh}, \bibinfo{person}{Joshua Yi~Min Tung},
  \bibinfo{person}{Jasmine Chiat~Ling Ong}, {and} \bibinfo{person}{Daniel
  Shu~Wei Ting}.} \bibinfo{year}{2024}\natexlab{}.
\newblock \showarticletitle{Development and Testing of Retrieval Augmented
  Generation in Large Language Models--A Case Study Report}.
\newblock \bibinfo{journal}{\emph{arXiv preprint arXiv:2402.01733}}
  (\bibinfo{year}{2024}).
\newblock


\bibitem[Klammt et~al\mbox{.}(2012)]%
        {Klammt2012}
\bibfield{author}{\bibinfo{person}{Christian Klammt},
  \bibinfo{person}{Innokentiy Maslennikov}, \bibinfo{person}{Monika Bayrhuber},
  \bibinfo{person}{Cédric Eichmann}, \bibinfo{person}{Navratna Vajpai},
  \bibinfo{person}{Ellis Jeremy~Chua Chiu}, \bibinfo{person}{Katherine~Y
  Blain}, \bibinfo{person}{Luis Esquivies}, \bibinfo{person}{June Hyun~Jung
  Kwon}, \bibinfo{person}{Bartosz Balana}, \bibinfo{person}{Ursula Pieper},
  \bibinfo{person}{Andrej Sali}, \bibinfo{person}{Paul~A Slesinger},
  \bibinfo{person}{Witek Kwiatkowski}, \bibinfo{person}{Roland Riek}, {and}
  \bibinfo{person}{Senyon Choe}.} \bibinfo{year}{2012}\natexlab{}.
\newblock \showarticletitle{Facile backbone structure determination of human
  membrane proteins by NMR spectroscopy}.
\newblock \bibinfo{journal}{\emph{Nature Methods}} \bibinfo{volume}{9},
  \bibinfo{number}{8} (\bibinfo{date}{May} \bibinfo{year}{2012}),
  \bibinfo{pages}{834–839}.
\newblock
\urldef\tempurl%
\url{https://doi.org/10.1038/nmeth.2033}
\showDOI{\tempurl}


\bibitem[Krishna et~al\mbox{.}(2024)]%
        {Krishna2024}
\bibfield{author}{\bibinfo{person}{Rohith Krishna}, \bibinfo{person}{Jue Wang},
  \bibinfo{person}{Woody Ahern}, \bibinfo{person}{Pascal Sturmfels},
  \bibinfo{person}{Preetham Venkatesh}, \bibinfo{person}{Indrek Kalvet},
  \bibinfo{person}{Gyu~Rie Lee}, \bibinfo{person}{Felix~S. Morey-Burrows},
  \bibinfo{person}{Ivan Anishchenko}, \bibinfo{person}{Ian~R. Humphreys},
  \bibinfo{person}{Ryan McHugh}, \bibinfo{person}{Dionne Vafeados},
  \bibinfo{person}{Xinting Li}, \bibinfo{person}{George~A. Sutherland},
  \bibinfo{person}{Andrew Hitchcock}, \bibinfo{person}{C.~Neil Hunter},
  \bibinfo{person}{Alex Kang}, \bibinfo{person}{Evans Brackenbrough},
  \bibinfo{person}{Asim~K. Bera}, \bibinfo{person}{Minkyung Baek},
  \bibinfo{person}{Frank DiMaio}, {and} \bibinfo{person}{David Baker}.}
  \bibinfo{year}{2024}\natexlab{}.
\newblock \showarticletitle{Generalized biomolecular modeling and design with
  RoseTTAFold All-Atom}.
\newblock \bibinfo{journal}{\emph{Science}} \bibinfo{volume}{384},
  \bibinfo{number}{6693} (\bibinfo{date}{April} \bibinfo{year}{2024}).
\newblock
\urldef\tempurl%
\url{https://doi.org/10.1126/science.adl2528}
\showDOI{\tempurl}


\bibitem[Lewis et~al\mbox{.}(2020)]%
        {lewis2020retrieval}
\bibfield{author}{\bibinfo{person}{Patrick Lewis}, \bibinfo{person}{Ethan
  Perez}, \bibinfo{person}{Aleksandra Piktus}, \bibinfo{person}{Fabio Petroni},
  \bibinfo{person}{Vladimir Karpukhin}, \bibinfo{person}{Naman Goyal},
  \bibinfo{person}{Heinrich K{\"u}ttler}, \bibinfo{person}{Mike Lewis},
  \bibinfo{person}{Wen-tau Yih}, \bibinfo{person}{Tim Rockt{\"a}schel},
  {et~al\mbox{.}}} \bibinfo{year}{2020}\natexlab{}.
\newblock \showarticletitle{Retrieval-augmented generation for
  knowledge-intensive nlp tasks}.
\newblock \bibinfo{journal}{\emph{Advances in Neural Information Processing
  Systems}}  \bibinfo{volume}{33} (\bibinfo{year}{2020}),
  \bibinfo{pages}{9459--9474}.
\newblock


\bibitem[Li et~al\mbox{.}(2013)]%
        {Li2013}
\bibfield{author}{\bibinfo{person}{Yanjun Li}, \bibinfo{person}{Yuanbo Zhao},
  \bibinfo{person}{Jia Hu}, \bibinfo{person}{Juan Xiao},
  \bibinfo{person}{Liujing Qu}, \bibinfo{person}{Zhenda Wang},
  \bibinfo{person}{Dalong Ma}, {and} \bibinfo{person}{Yingyu Chen}.}
  \bibinfo{year}{2013}\natexlab{}.
\newblock \showarticletitle{A novel ER-localized transmembrane protein, EMC6,
  interacts with RAB5A and regulates cell autophagy}.
\newblock \bibinfo{journal}{\emph{Autophagy}} \bibinfo{volume}{9},
  \bibinfo{number}{2} (\bibinfo{date}{February} \bibinfo{year}{2013}),
  \bibinfo{pages}{150–163}.
\newblock
\urldef\tempurl%
\url{https://doi.org/10.4161/auto.22742}
\showDOI{\tempurl}


\bibitem[Lin et~al\mbox{.}(2023)]%
        {Lin2023}
\bibfield{author}{\bibinfo{person}{Zeming Lin}, \bibinfo{person}{Halil Akin},
  \bibinfo{person}{Roshan Rao}, \bibinfo{person}{Brian Hie},
  \bibinfo{person}{Zhongkai Zhu}, \bibinfo{person}{Wenting Lu},
  \bibinfo{person}{Nikita Smetanin}, \bibinfo{person}{Robert Verkuil},
  \bibinfo{person}{Ori Kabeli}, \bibinfo{person}{Yaniv Shmueli},
  \bibinfo{person}{Allan Dos Santos~Costa}, \bibinfo{person}{Maryam
  Fazel-Zarandi}, \bibinfo{person}{Tom Sercu}, \bibinfo{person}{Salvatore
  Candido}, {and} \bibinfo{person}{Alexander Rives}.}
  \bibinfo{year}{2023}\natexlab{}.
\newblock \showarticletitle{Evolutionary-scale prediction of atomic-level
  protein structure with a language model}.
\newblock \bibinfo{journal}{\emph{Science}} \bibinfo{volume}{379},
  \bibinfo{number}{6637} (\bibinfo{date}{March} \bibinfo{year}{2023}),
  \bibinfo{pages}{1123--1130}.
\newblock


\bibitem[Liu et~al\mbox{.}(2013)]%
        {Liu2013}
\bibfield{author}{\bibinfo{person}{Jun Liu}, \bibinfo{person}{Fei Li}, {and}
  \bibinfo{person}{Sharon Rozovsky}.} \bibinfo{year}{2013}\natexlab{}.
\newblock \showarticletitle{The Intrinsically Disordered Membrane Protein
  Selenoprotein S Is a Reductase in Vitro}.
\newblock \bibinfo{journal}{\emph{Biochemistry}} \bibinfo{volume}{52},
  \bibinfo{number}{18} (\bibinfo{date}{April} \bibinfo{year}{2013}),
  \bibinfo{pages}{3051–3061}.
\newblock
\urldef\tempurl%
\url{https://doi.org/10.1021/bi4001358}
\showDOI{\tempurl}


\bibitem[MacLennan and Kranias(2003)]%
        {MacLennan2003}
\bibfield{author}{\bibinfo{person}{David~H. MacLennan} {and}
  \bibinfo{person}{Evangelia~G. Kranias}.} \bibinfo{year}{2003}\natexlab{}.
\newblock \showarticletitle{Phospholamban: a crucial regulator of cardiac
  contractility}.
\newblock \bibinfo{journal}{\emph{Nature Reviews Molecular Cell Biology}}
  \bibinfo{volume}{4}, \bibinfo{number}{7} (\bibinfo{date}{July}
  \bibinfo{year}{2003}), \bibinfo{pages}{566–577}.
\newblock
\urldef\tempurl%
\url{https://doi.org/10.1038/nrm1151}
\showDOI{\tempurl}


\bibitem[Maeda(2000)]%
        {Maeda2000}
\bibfield{author}{\bibinfo{person}{Y. Maeda}.} \bibinfo{year}{2000}\natexlab{}.
\newblock \showarticletitle{Human dolichol-phosphate-mannose synthase consists
  of three subunits, DPM1, DPM2 and DPM3}.
\newblock \bibinfo{journal}{\emph{The EMBO Journal}} \bibinfo{volume}{19},
  \bibinfo{number}{11} (\bibinfo{date}{June} \bibinfo{year}{2000}),
  \bibinfo{pages}{2475–2482}.
\newblock
\urldef\tempurl%
\url{https://doi.org/10.1093/emboj/19.11.2475}
\showDOI{\tempurl}


\bibitem[Makarewich et~al\mbox{.}(2018)]%
        {Makarewich2018}
\bibfield{author}{\bibinfo{person}{Catherine~A Makarewich},
  \bibinfo{person}{Amir~Z Munir}, \bibinfo{person}{Gabriele~G Schiattarella},
  \bibinfo{person}{Svetlana Bezprozvannaya}, \bibinfo{person}{Olga~N
  Raguimova}, \bibinfo{person}{Ellen~E Cho}, \bibinfo{person}{Alexander~H
  Vidal}, \bibinfo{person}{Seth~L Robia}, \bibinfo{person}{Rhonda Bassel-Duby},
  {and} \bibinfo{person}{Eric~N Olson}.} \bibinfo{year}{2018}\natexlab{}.
\newblock \showarticletitle{The DWORF micropeptide enhances contractility and
  prevents heart failure in a mouse model of dilated cardiomyopathy}.
\newblock \bibinfo{journal}{\emph{eLife}}  \bibinfo{volume}{7}
  (\bibinfo{date}{October} \bibinfo{year}{2018}).
\newblock
\urldef\tempurl%
\url{https://doi.org/10.7554/elife.38319}
\showDOI{\tempurl}


\bibitem[McInnes et~al\mbox{.}(2018)]%
        {McInnes2018}
\bibfield{author}{\bibinfo{person}{Leland McInnes}, \bibinfo{person}{John
  Healy}, \bibinfo{person}{Nathaniel Saul}, {and} \bibinfo{person}{Lukas
  Großberger}.} \bibinfo{year}{2018}\natexlab{}.
\newblock \showarticletitle{UMAP: Uniform Manifold Approximation and
  Projection}.
\newblock \bibinfo{journal}{\emph{Journal of Open Source Software}}
  \bibinfo{volume}{3}, \bibinfo{number}{29} (\bibinfo{date}{September}
  \bibinfo{year}{2018}), \bibinfo{pages}{861}.
\newblock
\urldef\tempurl%
\url{https://doi.org/10.21105/joss.00861}
\showDOI{\tempurl}


\bibitem[Rapoport(2008)]%
        {Rapoport2008}
\bibfield{author}{\bibinfo{person}{Tom~A. Rapoport}.}
  \bibinfo{year}{2008}\natexlab{}.
\newblock \showarticletitle{Protein transport across the endoplasmic reticulum
  membrane: Delivered on 8 July 2007 at the 32nd FEBS Congress in Vienna,
  Austria}.
\newblock \bibinfo{journal}{\emph{The FEBS Journal}} \bibinfo{volume}{275},
  \bibinfo{number}{18} (\bibinfo{date}{August} \bibinfo{year}{2008}),
  \bibinfo{pages}{4471–4478}.
\newblock
\urldef\tempurl%
\url{https://doi.org/10.1111/j.1742-4658.2008.06588.x}
\showDOI{\tempurl}


\bibitem[Salemi and Zamani(2024)]%
        {Salemi2024}
\bibfield{author}{\bibinfo{person}{Alireza Salemi} {and} \bibinfo{person}{Hamed
  Zamani}.} \bibinfo{year}{2024}\natexlab{}.
\newblock \showarticletitle{Evaluating Retrieval Quality in Retrieval-Augmented
  Generation}.
\newblock  (\bibinfo{year}{2024}).
\newblock
\urldef\tempurl%
\url{https://doi.org/10.48550/ARXIV.2404.13781}
\showDOI{\tempurl}


\bibitem[Scavone et~al\mbox{.}(2023)]%
        {Scavone2023}
\bibfield{author}{\bibinfo{person}{Francesco Scavone},
  \bibinfo{person}{Samantha C. Gumbin}, \bibinfo{person}{Paul A. Da Rosa},
  {and} \bibinfo{person}{Ron R. Kopito}.} \bibinfo{year}{2023}\natexlab{}.
\newblock \showarticletitle{RPL26/uL24 UFMylation is essential for
  ribosome-associated quality control at the endoplasmic reticulum}.
\newblock \bibinfo{journal}{\emph{Proceedings of the National Academy of
  Sciences}} \bibinfo{volume}{120}, \bibinfo{number}{16} (\bibinfo{date}{April}
  \bibinfo{year}{2023}).
\newblock
\urldef\tempurl%
\url{https://doi.org/10.1073/pnas.2220340120}
\showDOI{\tempurl}


\bibitem[Sgarbossa et~al\mbox{.}(2023)]%
        {Sgarbossa2023}
\bibfield{author}{\bibinfo{person}{Damiano Sgarbossa}, \bibinfo{person}{Umberto
  Lupo}, {and} \bibinfo{person}{Anne-Florence Bitbol}.}
  \bibinfo{year}{2023}\natexlab{}.
\newblock \showarticletitle{Generative power of a protein language model
  trained on multiple sequence alignments}.
\newblock \bibinfo{journal}{\emph{eLife}}  \bibinfo{volume}{12}
  (\bibinfo{date}{Feb.} \bibinfo{year}{2023}).
\newblock
\showISSN{2050-084X}
\urldef\tempurl%
\url{https://doi.org/10.7554/elife.79854}
\showDOI{\tempurl}


\bibitem[Sharma et~al\mbox{.}(2010)]%
        {Sharma2010}
\bibfield{author}{\bibinfo{person}{Parveen Sharma}, \bibinfo{person}{Vladimir
  Ignatchenko}, \bibinfo{person}{Kevin Grace}, \bibinfo{person}{Claudia
  Ursprung}, \bibinfo{person}{Thomas Kislinger}, {and}
  \bibinfo{person}{Anthony~O. Gramolini}.} \bibinfo{year}{2010}\natexlab{}.
\newblock \showarticletitle{Endoplasmic Reticulum Protein Targeting of
  Phospholamban: A Common Role for an N-Terminal Di-Arginine Motif in ER
  Retention?}
\newblock \bibinfo{journal}{\emph{PLoS ONE}} \bibinfo{volume}{5},
  \bibinfo{number}{7} (\bibinfo{date}{July} \bibinfo{year}{2010}),
  \bibinfo{pages}{e11496}.
\newblock
\urldef\tempurl%
\url{https://doi.org/10.1371/journal.pone.0011496}
\showDOI{\tempurl}


\bibitem[Shikano and Li(2003)]%
        {Shikano2003}
\bibfield{author}{\bibinfo{person}{Sojin Shikano} {and} \bibinfo{person}{Min
  Li}.} \bibinfo{year}{2003}\natexlab{}.
\newblock \showarticletitle{Membrane receptor trafficking: Evidence of proximal
  and distal zones conferred by two independent endoplasmic reticulum
  localization signals}.
\newblock \bibinfo{journal}{\emph{Proceedings of the National Academy of
  Sciences}} \bibinfo{volume}{100}, \bibinfo{number}{10} (\bibinfo{date}{April}
  \bibinfo{year}{2003}), \bibinfo{pages}{5783–5788}.
\newblock
\urldef\tempurl%
\url{https://doi.org/10.1073/pnas.1031748100}
\showDOI{\tempurl}


\bibitem[Silvente-Poirot and Poirot(2012)]%
        {SilventePoirot2012}
\bibfield{author}{\bibinfo{person}{Sandrine Silvente-Poirot} {and}
  \bibinfo{person}{Marc Poirot}.} \bibinfo{year}{2012}\natexlab{}.
\newblock \showarticletitle{Cholesterol epoxide hydrolase and cancer}.
\newblock \bibinfo{journal}{\emph{Current Opinion in Pharmacology}}
  \bibinfo{volume}{12}, \bibinfo{number}{6} (\bibinfo{date}{December}
  \bibinfo{year}{2012}), \bibinfo{pages}{696–703}.
\newblock
\urldef\tempurl%
\url{https://doi.org/10.1016/j.coph.2012.07.007}
\showDOI{\tempurl}


\bibitem[Sun and Kekenes-Huskey(2023)]%
        {Sun2023}
\bibfield{author}{\bibinfo{person}{Bin Sun} {and} \bibinfo{person}{Peter~M.
  Kekenes-Huskey}.} \bibinfo{year}{2023}\natexlab{}.
\newblock \showarticletitle{Myofilament-associated proteins with intrinsic
  disorder (MAPIDs) and their resolution by computational modeling}.
\newblock \bibinfo{journal}{\emph{Quarterly Reviews of Biophysics}}
  \bibinfo{volume}{56} (\bibinfo{year}{2023}).
\newblock
\showISSN{1469-8994}
\urldef\tempurl%
\url{https://doi.org/10.1017/s003358352300001x}
\showDOI{\tempurl}


\bibitem[Traaseth et~al\mbox{.}(2008)]%
        {Traaseth2008}
\bibfield{author}{\bibinfo{person}{Nathaniel~J Traaseth},
  \bibinfo{person}{Kim~N Ha}, \bibinfo{person}{Raffaello Verardi},
  \bibinfo{person}{Lei Shi}, \bibinfo{person}{Jarrod~J Buffy},
  \bibinfo{person}{Larry~R Masterson}, {and} \bibinfo{person}{Gianluigi
  Veglia}.} \bibinfo{year}{2008}\natexlab{}.
\newblock \showarticletitle{{Structural and Dynamic Basis of Phospholamban and
  Sarcolipin Inhibition of Ca 2+-ATPase †}}.
\newblock \bibinfo{journal}{\emph{Biochemistry}} \bibinfo{volume}{47},
  \bibinfo{number}{1} (\bibinfo{year}{2008}), \bibinfo{pages}{3--13}.
\newblock
\urldef\tempurl%
\url{http://pubs.acs.org/doi/abs/10.1021/bi701668v
  papers2://publication/doi/10.1021/bi701668v}
\showURL{%
\tempurl}


\bibitem[Wang et~al\mbox{.}(2024)]%
        {Wang2024}
\bibfield{author}{\bibinfo{person}{Chao Wang}, \bibinfo{person}{Hehe Fan},
  \bibinfo{person}{Ruijie Quan}, {and} \bibinfo{person}{Yi Yang}.}
  \bibinfo{year}{2024}\natexlab{}.
\newblock \showarticletitle{ProtChatGPT: Towards Understanding Proteins with
  Large Language Models}.
\newblock  (\bibinfo{year}{2024}).
\newblock
\urldef\tempurl%
\url{https://doi.org/10.48550/ARXIV.2402.09649}
\showDOI{\tempurl}


\bibitem[Wasim et~al\mbox{.}(2024)]%
        {Wasim2024}
\bibfield{author}{\bibinfo{person}{Abdul Wasim}, \bibinfo{person}{Ushasi
  Pramanik}, \bibinfo{person}{Anirban Das}, \bibinfo{person}{Pikaso Latua},
  \bibinfo{person}{Jai~S. Rudra}, {and} \bibinfo{person}{Jagannath Mondal}.}
  \bibinfo{year}{2024}\natexlab{}.
\newblock \showarticletitle{Harnessing Transformers to Generate Protein
  Sequences Prone to Liquid Liquid Phase Separation}.
\newblock  (\bibinfo{date}{March} \bibinfo{year}{2024}).
\newblock
\urldef\tempurl%
\url{https://doi.org/10.1101/2024.03.02.583105}
\showDOI{\tempurl}


\bibitem[Wei et~al\mbox{.}(2022)]%
        {Wei2022}
\bibfield{author}{\bibinfo{person}{Jason Wei}, \bibinfo{person}{Xuezhi Wang},
  \bibinfo{person}{Dale Schuurmans}, \bibinfo{person}{Maarten Bosma},
  \bibinfo{person}{Brian Ichter}, \bibinfo{person}{Fei Xia},
  \bibinfo{person}{Ed Chi}, \bibinfo{person}{Quoc Le}, {and}
  \bibinfo{person}{Denny Zhou}.} \bibinfo{year}{2022}\natexlab{}.
\newblock \showarticletitle{Chain-of-Thought Prompting Elicits Reasoning in
  Large Language Models}.
\newblock  (\bibinfo{year}{2022}).
\newblock
\urldef\tempurl%
\url{https://doi.org/10.48550/ARXIV.2201.11903}
\showDOI{\tempurl}


\bibitem[Wu et~al\mbox{.}(2023)]%
        {Wu2023}
\bibfield{author}{\bibinfo{person}{Fang Wu}, \bibinfo{person}{Lirong Wu},
  \bibinfo{person}{Dragomir Radev}, \bibinfo{person}{Jinbo Xu}, {and}
  \bibinfo{person}{Stan~Z Li}.} \bibinfo{year}{2023}\natexlab{}.
\newblock \showarticletitle{Integration of pre-trained protein language models
  into geometric deep learning networks}.
\newblock \bibinfo{journal}{\emph{Commun. Biol.}} \bibinfo{volume}{6},
  \bibinfo{number}{1} (\bibinfo{date}{Aug.} \bibinfo{year}{2023}),
  \bibinfo{pages}{876}.
\newblock


\bibitem[Zhao et~al\mbox{.}(2021)]%
        {Zhao2021}
\bibfield{author}{\bibinfo{person}{Mingzhe Zhao}, \bibinfo{person}{Jingsong
  Ma}, \bibinfo{person}{Mo Li}, \bibinfo{person}{Yingtian Zhang},
  \bibinfo{person}{Bixuan Jiang}, \bibinfo{person}{Xianglong Zhao},
  \bibinfo{person}{Cong Huai}, \bibinfo{person}{Lu Shen}, \bibinfo{person}{Na
  Zhang}, \bibinfo{person}{Lin He}, {and} \bibinfo{person}{Shengying Qin}.}
  \bibinfo{year}{2021}\natexlab{}.
\newblock \showarticletitle{Cytochrome P450 Enzymes and Drug Metabolism in
  Humans}.
\newblock \bibinfo{journal}{\emph{International Journal of Molecular Sciences}}
  \bibinfo{volume}{22}, \bibinfo{number}{23} (\bibinfo{date}{November}
  \bibinfo{year}{2021}), \bibinfo{pages}{12808}.
\newblock
\urldef\tempurl%
\url{https://doi.org/10.3390/ijms222312808}
\showDOI{\tempurl}


\end{thebibliography}

\appendix

\clearpage
\newpage
\section{Supplementary}
\subsection{Sample prompts}
\label{supp:prompts}
\lbi
\item Ex 1.
\textsc{The sequence 'xxxx' with numerical attributes x,y,z is predicted to bind SERCA. Based on this information, would the sequence 'yyy' with numerical attributes u,v,w bind SERCA and why?}
\lei

\subsection{Equations}
\label{equations}
We define the Shannon entropy as 
\begin{equation}
\label{eqn:entropy}
E(p) = -p\;\log 2 (p) + (1-p)\;\log 2 (1-p)
\end{equation}
where $p$ is the percentage of points with output=1.

The centroid hydrophobicity quantifies the clustering of hydrophobic residues around the midpoint of a sequence, under the assumption that sequences rich in hydrophobic residues will embed in the ER bilayer. This is calculated using the equation:

\begin{equation}
\label{eqn:hydrophobicity}
c_h = \frac{1}{2\delta}\sum_{j=i-\delta}^{i+\delta} h_j
\end{equation}

\noindent Here, $i$ represents the position of the central residue, and $h_j = 1$ if the residue at position $j$ is hydrophobic and $0$ otherwise.

Similarly, the charge density is evaluated by counting the number of D (Aspartic acid), E (Glutamic acid), K (Lysine), and R (Arginine) residues, and then dividing this sum by the total number of residues in the sequence.

%
\begin{table}[h]
    \centering
    \caption{Parameters used \label{tbl:params}. Refer also to \sect{yaml}}
    \begin{tabular}{ll}
        \hline
        \textbf{Parameter} & \textbf{Description} \\
        \hline
        Token length & Length of tokens used for processing \\
        Chunks & Size of data segments for processing \\
        Overlap & Overlapping amount between consecutive chunks \\
        Numbers of clusters & Number of clusters in the data \\
        nInputOutputPairs & Number of input-output pairs for training \\ \bottomrule
    \end{tabular}
\end{table}

\subsection{Inputs}
\subsubsection{Yaml parameter files}
\label{yaml}
Below is an example yaml file used for this study.
\begin{verbatim}
modelType: llama3:70b
embedding_model: thenlper/gte-small
chunk_size: 100
chunk_overlap: 0
dbPath: "/data/pkekeneshuskey/db/huggf_db_seq"
srcPath: "../sequences/training.txt"
testPath: "../sequences/100"
chunkByLine: True
testing: False
manual: False
nClusters: 5
nPromptsWanted: 10
appendFeatures: True
prompt:  >
  <s> [INST]
  Each line in the provided context contains one input, a sequence of A,C,D,E,F,G,H,I,K,L,M,N,P,Q,R,S,T,V,W,Y,0, and 1, several numbers, as well as one output digit.
  [/INST] </s>
  [INST] Question: {question}
  Context: {context}
  Answer: [/INST]
query:
  - Will be provided.
\end{verbatim}

\subsubsection{Testing with controls} 
\label{supp:test}
\label{supp:controls}
Prompt for testing controls:

\begin{verbatim}
I have a list of input/output pairs, where the input includes an amino acid sequence, a score,
a length, and a hydrophobicity score, and the output is 1 or 0 if it binds a target protein.
We also observed that output 1 sequences have adjacent RR residues in the FASTA input. 
Input: RNVYVFLATSGTLAGIMGMRFYHSGKFMPAGLIAGASLLMVAKVGVSMFNRPH0,0.064,54,0.1  Output: 0;
Input: MAAVVAKREGPPFISEAAVRGNAAVLDYCRTSVSALSGATGGILGLTGLY1,0.065,51,0.1  Output: 0;
Input: GVIATIAFLMINAVSNGQVRGDSYSEGCLGQTAARIWLFVGFMLAFGSLIASMWI0,0.072,56,0.1  Output: 1;
Input: MAAVVAKREGPPFISEAAVRGNAAVLDYCRTSVSALSGAT0,0.060,41,0.2  Output: 0;
Input: MEEASEGGGNDRVRNLQTEVEGVKNIMTQNVERILARGENLEHL1,0.043,45,0.3  Output: 1;
Input: MWRFYTEDSPGLKVGPVPVIVMSLLFIASVFML1,0.045,34,0.1  Output: 1;
Input: SGKFMPAGLIAGASLLMVAKVGVSMFNRPH1,0.056,31,0.1  Output: 0;
Input: SETEEDTSSSPHRIRSARQRRAPADEGHRPLS1,0.030,33,0.4  Output: 1;
Input: MSTGPTAATGSNRRLQQTQNQVDEVVDLMRVNVDKVLERDQKLSELDDRA0,0.051,51,0.3  Output: 1;
Input: IMTQNVERILARGENLEHLRNKTEDLEATSEHFKTTSQKVAKKFWWKNVKMIVL1,0.054,55,0.3  Output: 1;
Based on the provided inputs and outputs, predict the output of the following inputs. Please write only your predicted output as the sequence and integer on its own line, terminated with a semicolon, such as 'Sequence1 Output: 1;' or 'Sequence 1 Output: 0;' Make your best guess if you don't know:
MEKVQYLTRSAIRRASTIEMPQQARQKLQNLFINFCLILICLLLI,0.114,45,0.2 ;
MEKVQYLTRSAIRRASTIEMPQQARQKLQNLFINFCLILICLLLI,0.114,45,0.2 ;
LGILIVLSRRCRCKFNQQQRTGEPDEEEGTFRSSIRRLSTRRR,0.137,43,0.4 ;
RCMKDDSKGKSEEELSDLFRMFDKNADGYIDLDELKIMLQATGETITEDDIEELMK,0.130,56,0.4 ;
DDIYKAAVEQLTEEQKNEFKAAFDIFVLGAEDGCISTKELGKVMRMLGQNPTPEELQE,0.129,58,0.3 ;
GNGTVMGAELRHVLATLGEKMKEEEVEALMAGQEDSNGCINYEAFVKHIMSI,0.117,52,0.2 ;
KIDLSAIKIEFSKEQQDEFKEAFLLFDRTGDSKITLSQVGDVLRALGT,0.119,48,0.3 ;
TQAEVLRVLGKPRQEELNTKMMDFETFLPMLQHISKNKDTGTYEDFVEGLRVFDKE,0.122,56,0.3 ;
ELRHVLATLGERLTEDEVEKLMAGQEDSNGCINYEAFVKHIMSS,0.135,44,0.3 ;
AGIHETTYNSIMKCDIDIRKDLYANNVMSGGTTMYPGIADRMQKEITALAPSTMKIKIIAP,0.139,61,0.2 ;
SAGIHETTYNSIMKCDIDIRKDLYANNVMSGGTTMYPGIADRMQKEITALAPSTMKIKIIAP,0.135,62,0.2
\end{verbatim}

\begin{table}[h!]
\centering
\begin{tabular}{lccl}
\hline
\textbf{Src} & \textbf{Seq} & \textbf{Binding?} & \textbf{Protein} \\ \hline
P26678 & RRASTIEMPQQARQ & Y & PLN \\
--     &  RASTIEMPQQARQ & Y & PLN \\
O00168 & RTGEPDEEEGTFRS & N & FYKD1 \\
P63316 & KIMLQATGETITEDDIEELMKD & N & TNNC1 \\ 
P63316 & KAAVEQLTEEQKNEFKA & N & TNNC1 \\ 
P08590 & KLMAGQEDSNGCINYEAFVKH & N & MYL3 \\  
P08590 & KNKDTGTYEDFVEGLRV & N & MYL3 \\ 
P68133 & KDLYANNVMSGGTTMYPGIADRM & N & ACTA1 \\  
P68133 & RKDLYANNVMSGGTTMYPGIADRM & N & ACTA1 \\ \hline
\end{tabular}
\caption{\label{supp:tblnovel}
Phenotypic Sequences and Associated Proteins}
\end{table}

\section{Online Resources}
Source code in support of this study will be made available at 
\url{https://github.com/huskeypm/pkh-lab-analyses}.
A colab notebook demonstrating the RAG concept (\sect{meth:rag})is also provided.

\end{document}